\documentclass[11pt,a4paper]{article}
\usepackage{jcappub}
\usepackage{epsfig}

\title{The stacked ISW signal of rare superstructures in $\Lambda$CDM}

\author[a]{Samuel Flender,}
\author[a]{Shaun Hotchkiss}
\author[b]{and Seshadri Nadathur}
\affiliation[a]{Department of Physics, University of Helsinki and Helsinki Institute of Physics, \\
P.O. Box 64, FIN-00014, University of Helsinki, Finland}
\affiliation[b]{Fakult\"at f\"ur Physik, Universit\"at Bielefeld,\\
Postfach 100131, D-33501 Bielefeld, Germany} 

\emailAdd{seshadri@physik.uni-bielefeld.de}

\abstract{A detection of the stacked integrated Sachs-Wolfe (ISW) signal in the CMB of rare superstructures identified in the SDSS Luminous Red Galaxy catalogue has been reported at very high statistical significance. The magnitude of the observed signal has previously been argued to be more than $3\sigma$ larger than the theoretical $\Lambda$CDM expectation. However, this calculation was made in the linear approximation, and relied on assumptions that may potentially have caused the $\Lambda$CDM expectation to be underestimated. Here we update the theoretical model calculation and compare it with an analysis of ISW maps obtained from $N$-body simulations of a $\Lambda$CDM universe. The differences between model predictions and the map analyses are found to be small and cannot explain the discrepancy with observation, which remains at $>3\sigma$ significance. We discuss the cosmological significance of this anomaly and speculate on the potential of alternative models to explain it.}

\keywords{CMBR experiments, CMBR theory, integrated Sachs-Wolfe effect, cosmological simulations}
\arxivnumber{}

\begin{document}
\maketitle
\flushbottom

\newcommand{\muk}{\,\rm{{\mu}K}}
\newcommand{\Mpc}{\,\rm{Mpc}}
\newcommand{\mpc}{\,h^{-1}\,\rm{Mpc}}
\newcommand{\dg}{^{\circ}}
\newcommand{\lcdm}{$\Lambda$CDM~}
\newcommand{\fnl}{$f_\mathrm{NL}$~}
\newcommand{\gnl}{$g_\mathrm{NL}$~}
\newcommand{\taunl}{$\tau_\mathrm{NL}$~}
\newcommand{\beq}{\begin{equation}}
\newcommand{\eeq}{\end{equation}}


\section{Introduction}
\label{section:intro}

The late-time integrated Sachs-Wolfe (ISW) effect can be used as a cosmological probe that is sensitive to the dynamical effects of dark energy \cite{Crittenden:1995ak}. It is manifested as secondary anisotropies in the cosmic microwave background (CMB) radiation which are introduced as photons from the last scattering surface travel through time-evolving gravitational potentials \cite{Sachs:1967er}. Under the assumption of spatial flatness and at linear order, the late-time evolution of potentials only occurs in an accelerating Universe. The detection of the ISW effect at linear order is therefore an important independent confirmation of the effects of dark energy, and can be used to test the standard $\Lambda$ Cold Dark Matter ($\Lambda$CDM) cosmological model.

Unfortunately, the amplitude of the ISW effect is an order of magnitude smaller than the intrinsic fluctuations in the CMB. Therefore detection of the signal requires careful cross-correlation of the CMB with data from large-scale structure catalogues that trace the matter density distribution \cite{Afshordi:2004kz,Douspis:2008xv}. Several such studies have been performed using different tracers, \emph{e.g.} \cite{Boughn:2003yz,Afshordi:2003xu,Fosalba:2003ge,Nolta:2003uy,Padmanabhan:2004fy,Cabre:2006qm,Rassat:2006kq,Giannantonio:2006du,Raccanelli:2008bc,HernandezMonteagudo:2009fb,Francis:2009ps,Sawangwit:2009gd,LopezCorredoira:2010rr}.~A range of different results is obtained; for a summary, see \cite{Dupe:2010zs}.~The use of a combination of several large-scale structure data sets has been found to give overall significance of $\sim4\sigma$ for the observation of a cross-correlation \cite{Giannantonio:2008zi,Ho:2008bz,Giannantonio:2012aa}. Greater sensitivity is required to test whether the form of the cross power spectrum is as expected in $\Lambda$CDM, though there are intriguing hints that the amplitude of cross-correlation with the SDSS LRG catalogue may be larger than expected \cite{Giannantonio:2008zi,Ho:2008bz,Giannantonio:2012aa,Granett:2008dz}, as well from the WISE catalogue \cite{Goto:2012}.

An alternative approach to the detection of the ISW signal has been to focus on the contributions from individual superstructures alone \cite{Granett:2008ju, Granett:2008xb}. In this approach, the authors identified the most extreme over- and under-dense superstructures in distribution of LRGs in the SDSS DR6 MegaZ photometric catalogue, and then measured the CMB temperature along the lines of sight corresponding to the directions of the identified structures. By stacking the CMB images so obtained and filtering with a compensated top-hat filter, they reported detections of hot spots correlated with overdensities and cold spots with underdensities, which may be attributed to the ISW effect. The amplitude of the correlation was found to be significant at $>3\sigma$ each for hot and cold spots individually, and at $\sim4.4\sigma$ for both combined. This is the highest significance detection yet obtained using a single tracer population.

However, the size of this detected signal is difficult to understand within the $\Lambda$CDM model \cite{Hunt:2008wp,Inoue:2010rp}. A recent analysis in ref.~\cite{Nadathur:2011iu} showed that  the reported stacked signal is more than $3$ standard deviations larger than the maximum expectation in the $\Lambda$CDM model if the primordial density perturbations were Gaussian, even with optimistic assumptions. The analytic calculation in \cite{Nadathur:2011iu} was however fully linear\footnote{Ref.~\cite{Inoue:2010rp} accounted for some non-linear effects but used a less conservative method for calculating the expectation value of the signal. The qualitative conclusions were the same.} and based on certain theoretical assumptions including the spherical nature of the average density fluctuation, and therefore did not take into account a potential biasing effect arising from smearing of the LRG sample in the radial direction due to photometric redshift errors. The effects of these simplifications are described in greater detail in section \ref{section:limitations}. These simplifications might in principle have affected the predicted maximum value of the stacked signal. This left open the small possibility that the \emph{most} optimistic prediction from the $\Lambda$CDM model might still be compatible with observation. 

To test the quantitative effect of these simplifications on the calculation in ref.~\cite{Nadathur:2011iu} the predictions of the analytic model should be compared with results obtained from large $N$-body simulations. To our knowledge, there does not exist an $N$-body simulation of the ISW effect in $\Lambda$CDM which covers the same redshift range and angular footprint as the SDSS photometric survey. However, ref.~\cite{Cai:2010hx} provides two realizations of sky maps of the ISW temperature anisotropies, including non-linear contributions, calculated for $N$-body simulations based on the \lcdm model of the matter distribution in two different redshift intervals, corresponding to two slabs of depth $1\;h^{-1}$Gpc. By filtering and stacking images from these maps in the same manner as done originally in \cite{Granett:2008ju,Granett:2008xb} we estimate the size of the expected maximum possible signal in \lcdm without recourse to the assumptions made in the previous analytic treatment. Such a treatment gives an upper bound on the maximum possible ISW signal from superstructures and is unlikely to be achieved in any realistic observation. 

Comparison with this upper bound obtained from the simulated maps shows that the spherical model slightly overestimates the maximum possible signal at large redshifts and underestimates it at small redshifts. The numerical difference between the spherical model prediction and the theoretical maximum is however always small. In particular, it is not sufficient to explain the discrepancy between the observed stacked temperature signal and the $\Lambda$CDM expectation obtained from the spherical model.

Using the simulated ISW maps, we are also able to investigate the variation of the maximum possible signal-to-noise ratio with the number $N$ of stacked structures used and the radius $\theta_F$ of the applied compensated top-hat filter. We find that the stacked ISW signal of superstructures in a $\Lambda$CDM universe with Gaussian primordial perturbations is not observable for any choice of $N$ and $\theta_F$, unless the survey volume is much larger than the SDSS survey.~We conclude that the high significance detection reported in ref.~\cite{Granett:2008ju} is not compatible with the ISW effect in such a cosmology, and the conclusions of ref.~\cite{Nadathur:2011iu} are confirmed.

The structure of the paper is as follows. In section~\ref{section:ISW} we briefly introduce the necessary background theory and describe the methodology used in the observation of the stacked signal. We then turn to the analytic spherical model. The main features of the model and the improvements over ref.~\cite{Nadathur:2011iu} are described in section~\ref{section:expectation}, with several mathematical details relegated to Appendix~\ref{appendix}. We explain the assumptions made and potential limitations of the model in section~\ref{section:limitations}. Then in section~\ref{section:maps} we introduce the ISW maps produced from $N$-body simulations and describe our method of analysis. The results of this analysis and a comparison with the spherical model are presented in section~\ref{section:results}. Finally we conclude the paper with a discussion of the implications of the result in section~\ref{section:conclusions}.

\section{The ISW signal of superstructures}
\label{section:ISW}

\subsection{Theory}
\label{section:ISWtheory}

The ISW effect is a secondary temperature anisotropy that arises when photons from the CMB travel through time-evolving gravitational potentials. At late times in the standard flat cosmology, this time evolution arises due to the effect of the non-zero dark energy density $\Omega_{\Lambda}$. The temperature anisotropy $\Delta T(\hat n)$ thus induced is  \cite{Sachs:1967er}
\begin{equation}
\label{eq:ISW}
\Delta T (\hat n) = \frac{2}{c^3}\bar{T_0}\int_0^{r_{\rm L}}
\dot{\Phi}(r,z,\hat n) \,a\, dr,
\end{equation}
where $\bar{T_0}$ is the mean CMB temperature, $r_{\rm L}$ is the radial comoving distance to the last scattering surface, $a$ is the scale factor, $\dot{\Phi}$ is the time derivative of the gravitational potential and $c$ is the speed of light. 

The potential $\Phi$ is related to the density fluctuation in the conformal Newtonian gauge via the Poisson equation,
\begin{equation}
\label{eq:Poisson}
\nabla^2 \Phi(\mathbf{x},t)=4\pi G \bar{\rho}(t) a^2\delta(\mathbf{x}, t),
\end{equation}
where $\bar{\rho}(t)$ is the mean matter density of the universe and $\delta\equiv(\rho-\bar{\rho})/\bar{\rho}$ is the density contrast. Changing to Fourier space and taking the time derivative gives
\begin{equation}
\label{eq:phidotdeltadot}
\dot{\Phi}(\mathbf{k},t)=\frac{3}{2}\left(\frac{H_0}{k}\right)^2\Omega_{\rm m}
\left[\frac{\dot{a}}{a^2}\delta(\mathbf{k},t)-\frac{\dot\delta(\mathbf{k},t)}{a}\right].
\end{equation}

In ref.~\cite{Cai:2010hx}, the continuity equation $\dot\delta(k,t)+i\vec{k}\cdot\vec{p}(\vec{k},t)=0$ is used in eq.~\eqref{eq:phidotdeltadot} to produce the simulated ISW maps, where $\vec{p}(\vec{k},t)$ is the Fourier transform of the momentum density divided by the mean mass density, and can be obtained directly from the simulation. The analytic model of ref.~\cite{Nadathur:2011iu} and its slightly modified version introduced in section \ref{section:sphericalmodel} both use the linear growth approximation $\delta(\mathbf{k},t)=D(t)\delta(\mathbf{k},z=0)$, where $D(t)$ is the linear growth factor. Substituting this into eq. \eqref{eq:phidotdeltadot} yields:
\begin{equation}
\label{eq:phidot_FS}
\dot{\Phi}(\mathbf{k},z)=\frac{3}{2}\left(\frac{H_0}{k}\right)^2\Omega_{\rm m}
\frac{H(z)}{a}[1-\beta(z)]\delta(\mathbf{k},z),
\end{equation}
where $\beta(t)\equiv{d \ln D}/{d \ln a}$ denotes the linear growth rate. As this model also assumes spherical symmetry of the structures, eq.~\eqref{eq:phidot_FS} may be rewritten in real space as \cite{Nadathur:2011iu}
\begin{equation}
\label{eq:dotphispherical}
\dot{\Phi}(r,z)=\frac{3}{2}\Omega_{\rm m}H_0^2 G(z) F(r),
\end{equation}
where $G(z)=H(z)(1-\beta(z))D(z)/a$ is the ISW linear growth factor and 
\begin{equation}
\label{eq:F(r)}
F(r)=\int_0^r \frac{r^{\prime 2}}{r}\delta(r^\prime){\rm d}r^{\prime}+\int_r^\infty r'\delta(r'){\rm d}r' \,,
\end{equation}
with $\delta(r')$ evaluated at $z=0$.

\subsection{Observation of the stacked signal}
\label{section:observation}

The signal reported in ref.~\cite{Granett:2008ju} was obtained in the following way. The authors used a sample of 1.1 million luminous red galaxies (LRGs) with photometric redshifts in the range $0.4<z<0.75$, and a median redshift of $z\sim0.52$, from the SDSS release DR6 \cite{AdelmanMcCarthy:2007aa}, covering a footprint of $7500$ square degrees about the North Galactic pole. Within this sample, they identified over- and under-dense fluctuations in the galactic density field using the structure-finding algorithms {\small VOBOZ} \cite{Neyrinck:2004gj} and {\small ZOBOV} \cite{Neyrinck:2007gy} respectively. The structures thus found were ranked according to their density contrast; this is roughly correlated with both the likelihood of such a structure existing and the relative ISW signal produced by it. 

Given such a ranking, the $N$ top-ranked ``superclusters" and ``supervoids" in the sample were selected, and the CMB images in the directions along the lines of sight of these structures were stacked and averaged. The CMB dataset used was an inverse-variance weighted combination of the WMAP 5-year Q, V and W maps \cite{Hinshaw:2008kr} with Galactic foreground template maps subtracted and the KQ75 mask applied. The temperature in the stacked images was then averaged with a circular compensated top-hat filter of the form 
\begin{equation}
\label{eq:tophatfilter}
W(\theta;\theta_F)=\begin{cases}
	1,\,\,0 \le \theta \le \theta_F,\\
	-1,\,\, \theta_F <\theta \le \theta_{\rm out},
	\end{cases}
\end{equation}
where $\theta_{\rm out}=\sqrt{2}\theta_F$. The effect of the filter is thus to take the average temperature within an inner circle of radius $\theta_F$ and subtract from it the average temperature within an outer circular ring of equal area, in order to remove the effect of primary CMB fluctuations on scales larger than the filter width. The noise due to the primary CMB fluctuations was estimated by randomising the positions of the stacked directions on the actual CMB map, and by using model CMB maps smoothed to the WMAP resolution.

For the choice $N=50$ and $\theta_F=4\dg$, the result obtained was $\Delta T_{\rm c}=7.9\pm3.1\;\mu$K for the hot spots associated with superclusters, $\Delta T_{\rm v}=-11.3\pm3.1\;\mu$K for the cold spots associated with supervoids, and $\Delta T=9.6\pm2.2\;\mu$K for the two combined (clusters minus voids), \emph{i.e.}, a $4.4\sigma$ significant detection \cite{Granett:2008ju}. The variation in signal strength with change in $N$ and $\theta_F$ was also investigated and found to be maximum at $N=50$ and $\theta_F=4\dg$. It is noteworthy that the signal obtained is not frequency dependent \cite{Granett:2008ju} and is therefore unlikely to be caused by contamination by foregrounds or by extragalactic radio sources. 

A follow-up investigation by the same authors \cite{Granett:2008dz} using a template-fit analysis with an ISW template reconstructed from the LRG density field confirmed the existence of the high-significance correlation between the observed supercluster and supervoid locations and the hot and cold spots on the CMB. This study also did not find any evidence of foreground contamination.

\section{The analytic spherical model}
\label{section:sphericalmodel}

\subsection{Calculation of the expected signal}
\label{section:expectation}

The analytic spherical model for the stacked ISW signal of superstructures introduced in ref.~\cite{Nadathur:2011iu} is based on the statistics of peaks of a Gaussian-distributed random field. We assume that the matter density contrast $\delta$, smoothed on a length-scale $R_{\rm f}$ large enough that all fluctuations are within the linear regime, is such a random field. Then by using the corresponding matter power spectrum for the \lcdm cosmological model, one can calculate the predicted (comoving) number densities of different populations of maxima (and minima) in $\delta$. The differential number density of maxima is \cite{BBKS}
\beq
\label{eq:numberdensity}
\mathcal{N}_{\rm max}(\nu,x){\rm d}\nu{\rm d}x=\frac{e^{-\nu^2/2}}{(2\pi)^2R_\ast^3}f(x)\frac{{\rm exp}\left[-(x-\gamma\nu)^2/2(1-\gamma^2)\right]}{\left[2\pi(1-\gamma^2)\right]^{1/2}}{\rm d}\nu{\rm d}x\,,
\eeq
where $\nu=\delta/\sigma_0$, $x=-\nabla^2\delta/\sigma_2$, $\sigma_0$ and $\sigma_2$ are  members of a set of spectral moments of the power spectrum of the density field, and $\gamma$ and $R_\ast$ are spectral parameters related to these moments. Definitions for these quantities as well as the full closed-form expression for $f(x)$ are provided in Appendix \ref{appendix}. 

We then identify the points of maxima in $\delta$ with the overdense structures in the galaxy density field found in the SDSS LRG catalogue and points of minima with voids. The galaxy density field $\delta_{\rm g}$ provides a biased tracer of $\delta$. This could be modelled using a bias relationship $\delta_{\rm g}=b\delta$, but in order to obtain the most optimistic estimate of the maximum achievable ISW signal it is sufficient to assume simply that the largest fluctuations in $\delta_{\rm g}$ occur at the same locations as the largest fluctuations in $\delta$.

If a maximum (minimum) of given height (depth) $\delta_0=\nu\sigma_0$ and with a given central $x$ exists at a point ${\mathbf r}=0$, the mean profile of the density contrast $\delta(r;\nu,x)$ about that point can be calculated \cite{BBKS}. In general the structures about a point of extremum have a triaxial ellipsoid shape, but in the absence of any preferred direction one can average over all orientations of the axes and obtain a mean spherical profile about a peak~\cite{BBKS,Lahav:1991}:
\beq
\label{eq:profile}
\bar\delta_{\rm pk}(r;\nu,x) = \frac{1}{\sigma_0}\int_0^{\infty} \frac{k^2}{2\pi^2}\frac{\sin(kr)}{kr}P_{\rm f}(k)\left[\frac{\nu-x\gamma}{(1-\gamma^2)}+\frac{(x-\gamma\nu)R_\ast^2k^2}{3\gamma(1-\gamma^2)}\right]{\rm d}k\,,
\eeq
where $P_{\rm f}(k)$ denotes the matter power spectrum filtered on the scale $R_{\rm f}$ using a spherical Gaussian filter. As long as fluctuations remain within the linear regime, the symmetry between peaks and voids is maintained. This can be seen from the fact that eqs.~\eqref{eq:numberdensity} and \eqref{eq:profile} are unchanged under a change of sign of $\nu$ and $x$ for voids. 

We model the structures identified in the SDSS LRG catalogue as being due to spherically symmetric peaks or voids in the matter density field, with profiles of the form eq.~\eqref{eq:profile}. For a given choice of $\delta_0$ and $x$ we calculate the ISW temperature shift $\Delta T(\theta;\delta_0,x)$ caused by a structure with such a profile by placing the centre of the sphere at a redshift $z=0.52$ (corresponding to the median redshift of the LRG catalogue) and using eqs.~\eqref{eq:dotphispherical} and \eqref{eq:F(r)} to perform the ISW line-of-sight integral of eq.~\eqref{eq:ISW} for any photon incidence angle $\theta$ through the structure. To this temperature profile we then apply the same compensated top-hat filter of width $\theta_F$ (eq.~\eqref{eq:tophatfilter}) used in \cite{Granett:2008ju}. The resulting quantity $\Delta T(\delta_0,x;\theta_F)$ is the observationally relevant average temperature signal due to this structure.

The evaluation of $\mathcal{N}_{\rm max}(\nu,x)$ and $\Delta T(\delta_0,x;\theta_F)$ for the \lcdm model requires the specification of the cosmological parameters. We use the same parameter values as in ref.~\cite{Nadathur:2011iu}, which are the mean values obtained from a fit to WMAP 7-year \cite{Komatsu:2010fb} and SDSS DR7 data \cite{Abazajian:2008wr}. However, the precise choice is unimportant, since variation of parameter values within the ranges allowed by the concordance cosmology has a negligible effect on the spherical model prediction \cite{Nadathur:2011iu}. To calculate the matter power spectrum $P(k)$ at the requisite redshift we use {\small CAMB} \cite{Lewis:1999bs}.
 
To obtain an expectation value of the average signal caused by a sample of structures of different central density contrasts and radii\footnote{Note that the radius of a structure with given $\delta_0$ is determined by the normalized second derivative of the density field at the centre, $x=-\nabla^2\delta/\sigma_2$. Larger values of $x$ correspond to narrower profiles.} we now take a weighted average of $\Delta T(\delta_0,x;\theta_F)$ over the range of values of $\delta_0$ and $x$, with the weighting function chosen according to the relative number of such structures present in the sample of structures. Thus for a sample of the $N$ largest and most overdense structures contained within some comoving volume $V$, we obtain the expectation value
\beq
\label{eq:expectationDeltaT}
\langle\Delta T_{\rm pk}(N,\theta_F)\rangle = \frac{V}{N}\int_{\delta_0^c}^{1}\int_0^{x_{\rm cut}}\Delta T_{\rm pk}(\delta_0,x;\theta_F)\mathcal{N}_{\rm max}\sigma_0^{-1}{\rm d}\delta_0{\rm d}x\,,
\eeq
where $\delta_0^c$ and $x_{\rm cut}(\delta_0^c)$ are cutoff values chosen according to the contours of $\Delta T_{\rm pk}(\delta_0,x;\theta_F)$, such as to include only the $N$ peaks that produce the largest temperature shift. The corresponding calculation for the expected signal for voids simply reverses the sign of the result. The variance of the distribution of temperatures about this expectation value can be estimated by $(\langle\Delta T^2\rangle-\langle\Delta T\rangle^2)$. 

It is worth pointing out that the sample of structures over which the expectation value \eqref{eq:expectationDeltaT} is defined is not chosen to match the actual sample of structures in the observed LRG density field. Rather it includes only those structures within the specified volume which give the largest contribution to the integral, irrespective of whether the particular structure-identification algorithm used would find them or not. In other words, $\langle\Delta T(N,\theta_F)\rangle$ represents the expectation of the \emph{maximum} possible observable signal from spherical structures, and is expected to be achievable only if the structure-finding algorithm is $100\%$ efficient.

Note also that the procedure for calculation of $\langle\Delta T(N,\theta_F)\rangle$ outlined above is slightly different to that used previously in ref.~\cite{Nadathur:2011iu}. In that paper, the radius of a structure of given $\delta_0$ was fixed to be at the mean value of the distribution by setting $x$ to its expectation value $\langle x|\delta_0\rangle$. Here we account for the distribution in radii and restrict the sample to include only those structures which produce the largest temperature shift. Thus the present analysis is even more conservative than that in ref.~\cite{Nadathur:2011iu}.

For $N=50$ and $\theta_F=4^\circ$ and the survey volume $V$ chosen to match that of the SDSS DR6 catalogue, the spherical model gives a maximum expected average ISW signal of superstructures to be $\langle\Delta T\rangle=2.27\pm0.14\;\mu$K, slightly larger than in \cite{Nadathur:2011iu}. Nevertheless, comparing this value with the observed value of $\Delta T=9.6\pm2.2\;\mu$K, we see that the original estimate of a discrepancy between theory and observation at a level higher than $3\sigma$ remains. Indeed, the spherical model prediction for the maximum ISW signal remains at the level of the observational noise.

\subsection{Limitations of the model}
\label{section:limitations}

Although the spherical model described above aims to provide the \emph{most} optimistic estimate for the size of the stacked ISW signal from superstructures, it necessarily rests on certain simplifying assumptions. If one or more of these assumptions were to have a large effect on the result, it may be that the calculation underestimates the maximum expected ISW signal. In such a case the discrepancy between the $\Lambda$CDM model and the observation of ref.~\cite{Granett:2008ju} might be mitigated.

The first assumption of the model is to place the centres of all structures at the same comoving distance from the observer without accounting for any effects of the redshift distribution. In particular this does not allow for overlap of structures at different redshifts along the line of sight. It was originally argued \cite{Nadathur:2011iu} that since overlapping structures could both increase or decrease the total signal, depending on the sign of the density fluctuation, the net effect of this simplification would be small.
 
Another simplification is that the model uses the linear approximation for growth of structures. Although on such large scales and low redshifts one expects this to be valid in the mean, with non-linear dynamics providing only a small correction \cite{Cai:2010hx}, for the specific population of the most extreme fluctuations it may not hold. Such non-linear effects would be expected to break the symmetry of the predicted signal from voids and peaks.

The entire calculation is also based on the assumption of Gaussianity of the density perturbations. However, non-linear evolution under gravity itself induces a small skewness and kurtosis to the density field. Although negligible in the mean, this may be important for the extreme tails of the distribution investigated here. Any effect of skewness on the actual temperature distribution would also affect voids and peaks differently.

The assumption of sphericity in particular may lead to an underestimate of the maximum possible signal. Although the mean profiles over all structures are spherical, the specific population of structures that produce the maximum ISW signal is significantly biased towards structures elongated along the line-of-sight direction. Photometric redshift errors in the LRG sample cause a significant redshift smearing along the line-of-sight in the reconstructed 3D LRG density distribution. As a result of this smearing, the population of structures found by structure-finding algorithms will \emph{not} have isotropic deviations from sphericity. Instead it will be somewhat biased towards structures elongated along the line-of-sight direction, which are also those that produce the largest ISW temperature shift. This is despite the fact that the structure-finding algorithms themselves do not have any preferred directionality. As a result, despite the conservative analysis, it is in principle possible that the sample of superstructures actually chosen for analysis gives a larger average effect than the model predicts.

Finally, the spherical model considers all $N$ structures to be statistically independent. This is well justified for small values of $N$; however as $N$ increases this assumption becomes less valid. For large values $N$, the fact that $N-1$ extreme structures with specific values of $\delta$ and $x$ already exist within a given volume will necessarily alter the expected $\delta$ and $x$ for the $N^{\rm th}$ structure.

The net effect of all these assumptions on the spherical model prediction is not clear. Indeed the relative importance of each is expected to depend on the redshift interval under consideration.  What we wish to determine is whether the calculated value $\langle\Delta T(N,\theta_F)\rangle$ for a population of spherical structures can underestimate the true maximum ISW signal that might be observed on the sky with a perfect (and lucky) observation. To answer this in the next section we compare the model predictions with the maximal stacked signal obtained from simulated sky maps of the ISW effect. 

\begin{figure*}[t]
\begin{center}
\includegraphics[scale=0.28]{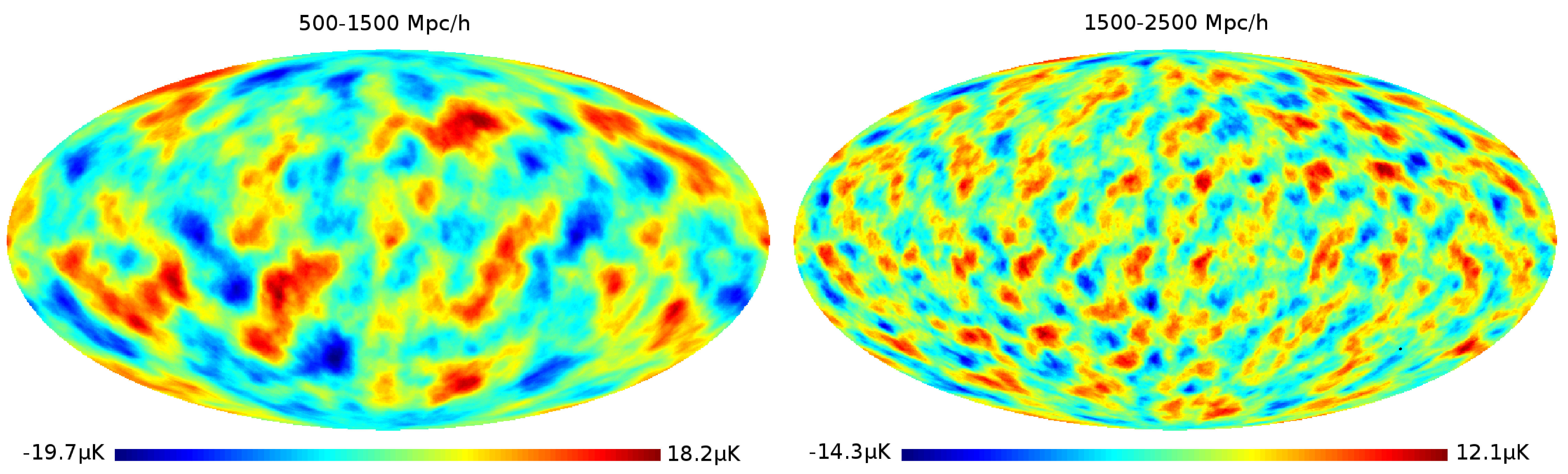}
\caption{\label{figure:mapsunfiltered} 
Sky maps of the ISW effect due to the matter density distribution in $N$-body simulations \cite{Cai:2010hx}. The method of generation of the maps is summarised in the text. Map A (left panel) is for the redshift range $0.17<z<0.57$, and Map B (right panel) for $0.57<z<1.08$. The corresponding comoving radial distance ranges are indicated in the figure. Note the asymmetry in the definition of the colour scale.}
\end{center}
\end{figure*}

\begin{figure*}[t]
\begin{center}
\includegraphics[scale=0.28]{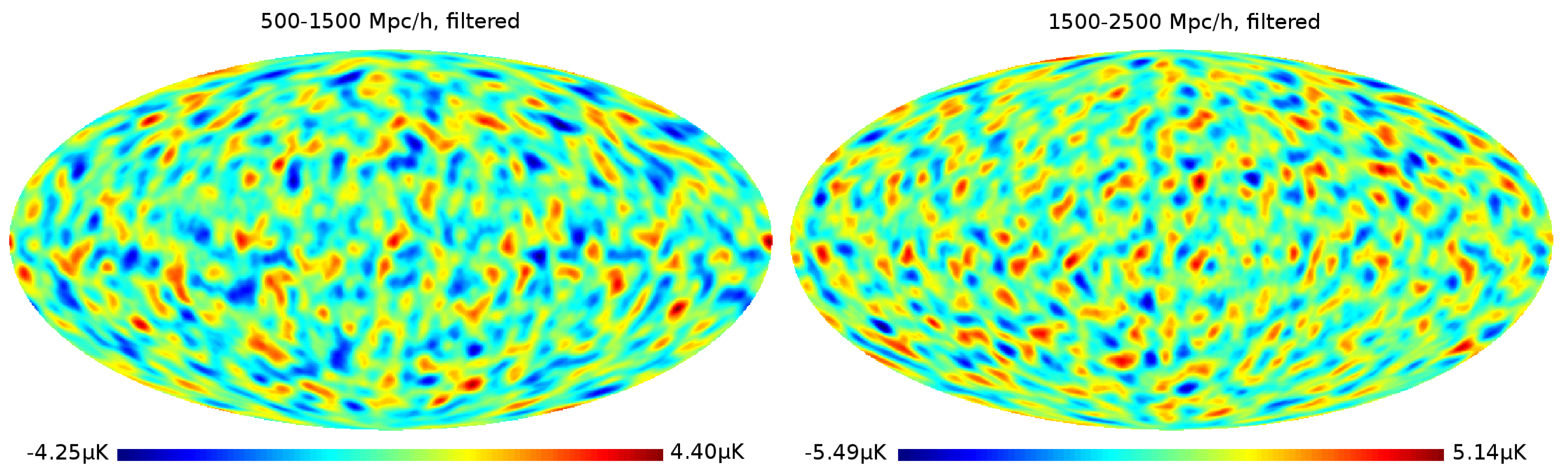}
\caption{\label{figure:mapsfiltered} 
The same maps as in figure~\ref{figure:mapsunfiltered}, but with the compensated top-hat filter $W(\theta;\theta_F)$ applied. The filter radius is $\theta_F=4\dg$ in this example.}
\end{center}
\end{figure*}

\section{Simulated sky maps}
\label{section:maps}

The ISW effect in a $\Lambda$CDM cosmology, including non-linear Rees-Sciama (RS) contributions \cite{Rees:1968zza}, has been studied using large $N$-body simulations in \cite{Cai:2010hx}. This study employed an $N$-body simulation of $2200^3$ particles in a box of side $1\;h^{-1}$Gpc, for a $\Lambda$CDM model with parameters $\Omega_\Lambda=0.74$, $\Omega_{\rm m}=0.26$, $\Omega_{\rm b}=0.044$, $\sigma_8=0.8$ and $H_0=71.5$ km s$^{-1}$ Mpc$^{-1}$. 

For our purposes here a brief summary of the method used to create the simulations will suffice. The initial density conditions were set at redshift $z=49$ and then evolved through 50 snapshots to redshift $z=0$, with the intervals between neighbouring simulation outputs corresponding to a radial comoving distance of $\sim100\;h^{-1}$Mpc. At each of these redshifts, the simulation output was used to construct $\dot\Phi$ according to eq.~\eqref{eq:phidotdeltadot}. By interpolating between these $\dot\Phi$ values, photon paths were then traced back through the simulation in order to calculate the cumulative $\Delta T$ effect via eq.~\eqref{eq:ISW} over depths up to the simulation depth of $1\;h^{-1}$Gpc. When the photon path exits the simulation box, it is mapped back to a location within the box using periodic boundary conditions. Finally the $\Delta T$ map thus created is visualised in spherical coordinates using HEALPix~\cite{Gorski:2004by}. 

In this way, full-sky maps of the ISW temperature shift including RS contributions are produced. Two of these maps, for the redshift intervals $0.17<z<0.57$ (corresponding to radial comoving distances of $500-1500\mpc$) and $0.57<z<1.08$ (corresponding to $1500-2500\mpc$), are shown in figure~\ref{figure:mapsunfiltered}.\footnote{We are grateful to Yan-Chuan Cai for providing us with these maps.} We refer to these as Map A and Map B respectively. Although some overlap is present, neither redshift range corresponds exactly to that of the SDSS LRG sample described in section~\ref{section:observation}, $0.4<z<0.75$. Therefore results obtained from these maps cannot be directly compared to observed values. However, they can be compared to the theoretical predictions for the same redshift intervals made using the spherical model of section~\ref{section:sphericalmodel}.

As can be seen in figure~\ref{figure:mapsunfiltered}, the magnitude of the temperature fluctuations are larger for Map A, which covers smaller redshifts. This is both because the effect of $\Lambda$ on the time evolution of $\Phi$ is larger at later times, and also because structures are more pronounced. It can also be seen that Map A has more power on larger scales. This is expected in any case due to the hierarchical growth of density perturbations; however, due to the finite box size and the periodic boundary conditions both maps also lack power and show non-Gaussianities on angular scales larger than the angle subtended by the $1\;h^{-1}$Gpc box size \cite{Cai:2010hx}. This scale is naturally smaller at larger comoving distances.

\begin{figure*}[t]
\begin{center}
\includegraphics[scale=0.4]{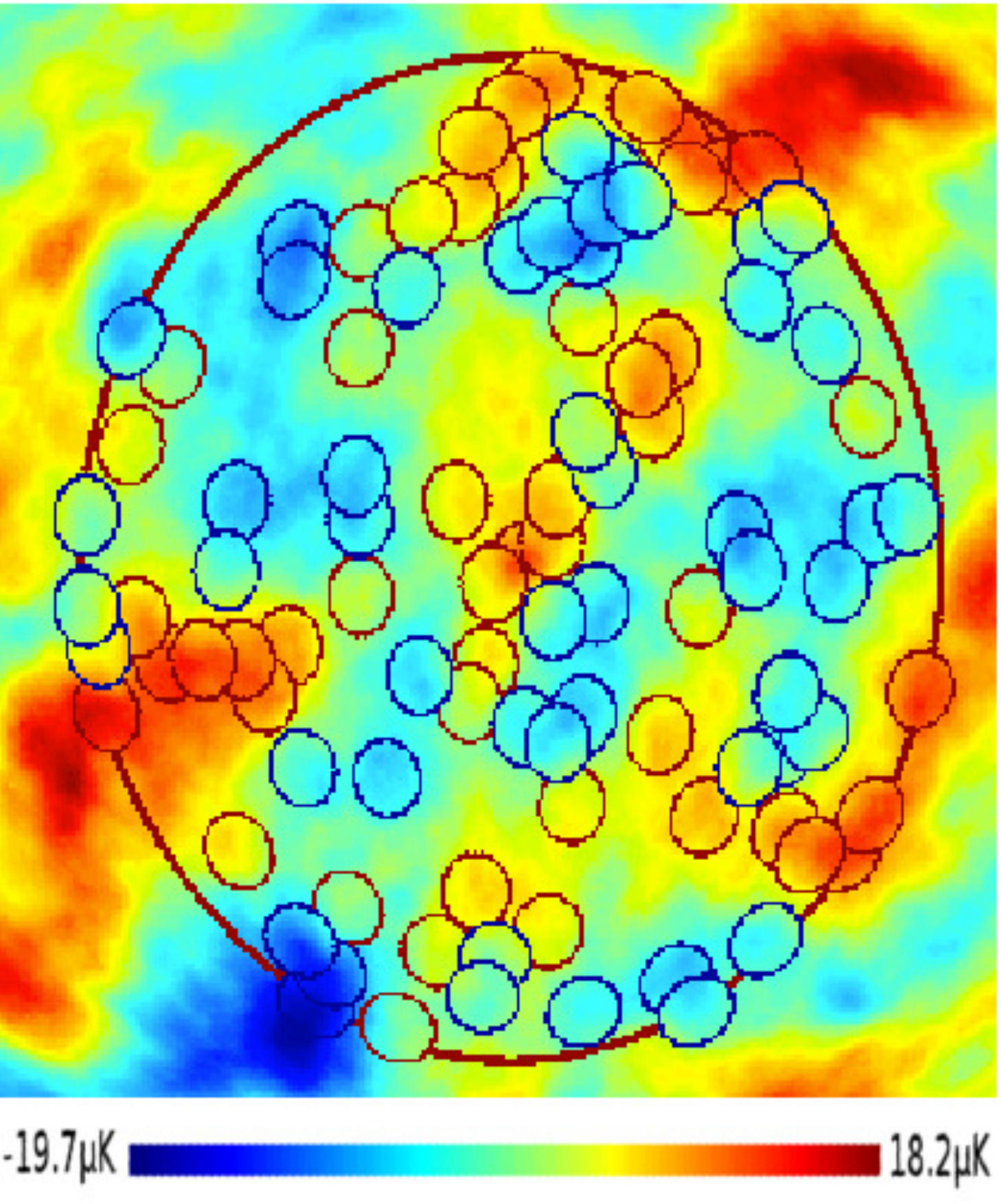}~~~~
\includegraphics[scale=0.4]{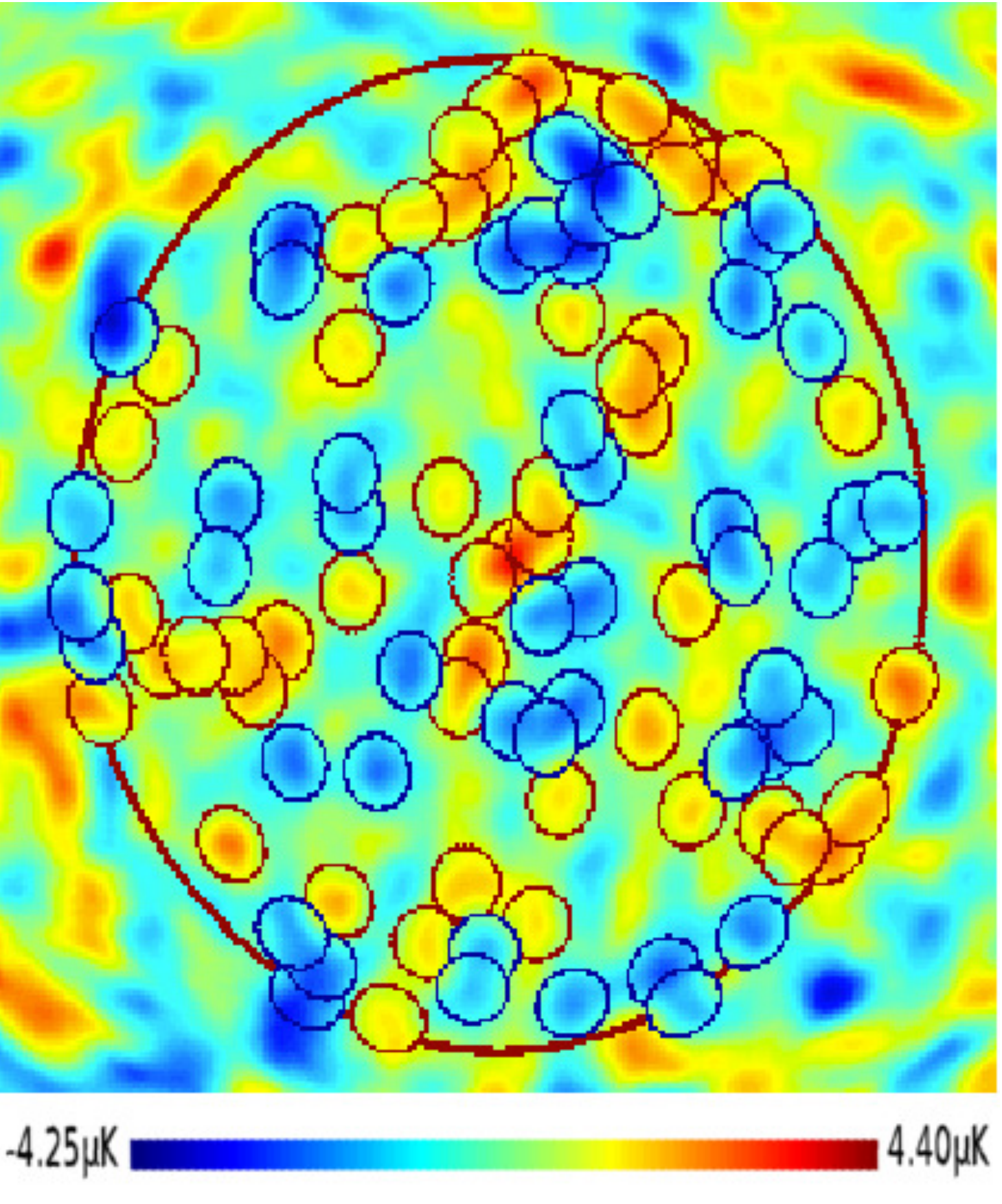}
\caption{\label{figure:selectionexample} 
Magnified section of Map A, illustrating the procedure for selecting the $50$ most extreme hot and cold pixels within a given window. The large circle defines the window area of $\sim8000$ square degrees. The smaller circular patches within are $4\dg$ circles centred on the selected pixels. Red circles indicate hot spots and blue circles cold spots. In the left-hand panel, the patches are shown superimposed on the unfiltered map, in the right-hand panel on the same area of the filtered map. Note that the actual selection is always performed on the filtered map.}
\end{center}
\end{figure*}

\subsection{Method of analysis}
\label{section:analysis}

We wish to use these two sky maps of the ISW effect to test the spherical model prediction for the maximum possible stacked signal of superstructures. To do this, we simply work backwards, by applying the methodology of ref.~\cite{Granett:2008ju} to the hottest and coldest spots that can be found in the given map, without any reference to actual structures observed in any galaxy survey. Clearly the value of $\Delta T$ thus obtained is not a realistic estimate of the expectation from an actual observation, but it does provide a robust upper bound to the observable stacked signal in $\Lambda$CDM.

First, we apply the compensated top-hat filter of eq.~\eqref{eq:tophatfilter} to each pixel of the original maps. For a choice of filter radius $\theta_F=4^\circ$ the resultant filtered maps can be seen in figure~\ref{figure:mapsfiltered} and compared with the equivalent unfiltered maps in figure~\ref{figure:mapsunfiltered}. 

We then select a circular window on the filtered map of angular size $\sim8000$ square degrees. The size of this window is chosen roughly to match the size of the SDSS DR9 footprint on the sky, and is somewhat larger than the footprint of the DR6 survey used in \cite{Granett:2008ju}.\footnote{Of course, the circular shape of the window does not correspond to the actual SDSS footprint, but the difference due to the geometry is small. We repeated the analysis with other  shapes of the window and found no significant change to the result.} Within this window, we select the $N$ hottest and $N$ coldest pixels. 

In order to ensure that not all pixels are chosen from the same structure on the map, we apply the following algorithm to limit overlap. For hot spots, we start by selecting the hottest pixel within the window after filtering on scale $\theta_F$. We then remove the circular region of radius $\theta_F$ about this direction from consideration and choose the next hottest pixel, before repeating the procedure to obtain $N$ hot spots. The same procedure applied to the coldest pixels gives $N$ cold spots. This ensures that the selected pixels are always separated by at least $\theta_F$, but some overlap of the circular regions over which the top-hat filter contributes to the pixel temperature is still allowed. Figure~\ref{figure:selectionexample} illustrates the result of one such selection process. Averaging the temperatures of these pixels gives the quantities $\Delta T_{\rm hot}$, $\Delta T_{\rm cold}$ and $\Delta T=(\Delta T_{\rm hot}-\Delta T_{\rm cold})/2$ as functions of $N$ and $\theta_F$ for the given location of the circular window on the sky.

\begin{figure}[t]
\begin{center}
\includegraphics[scale=0.5]{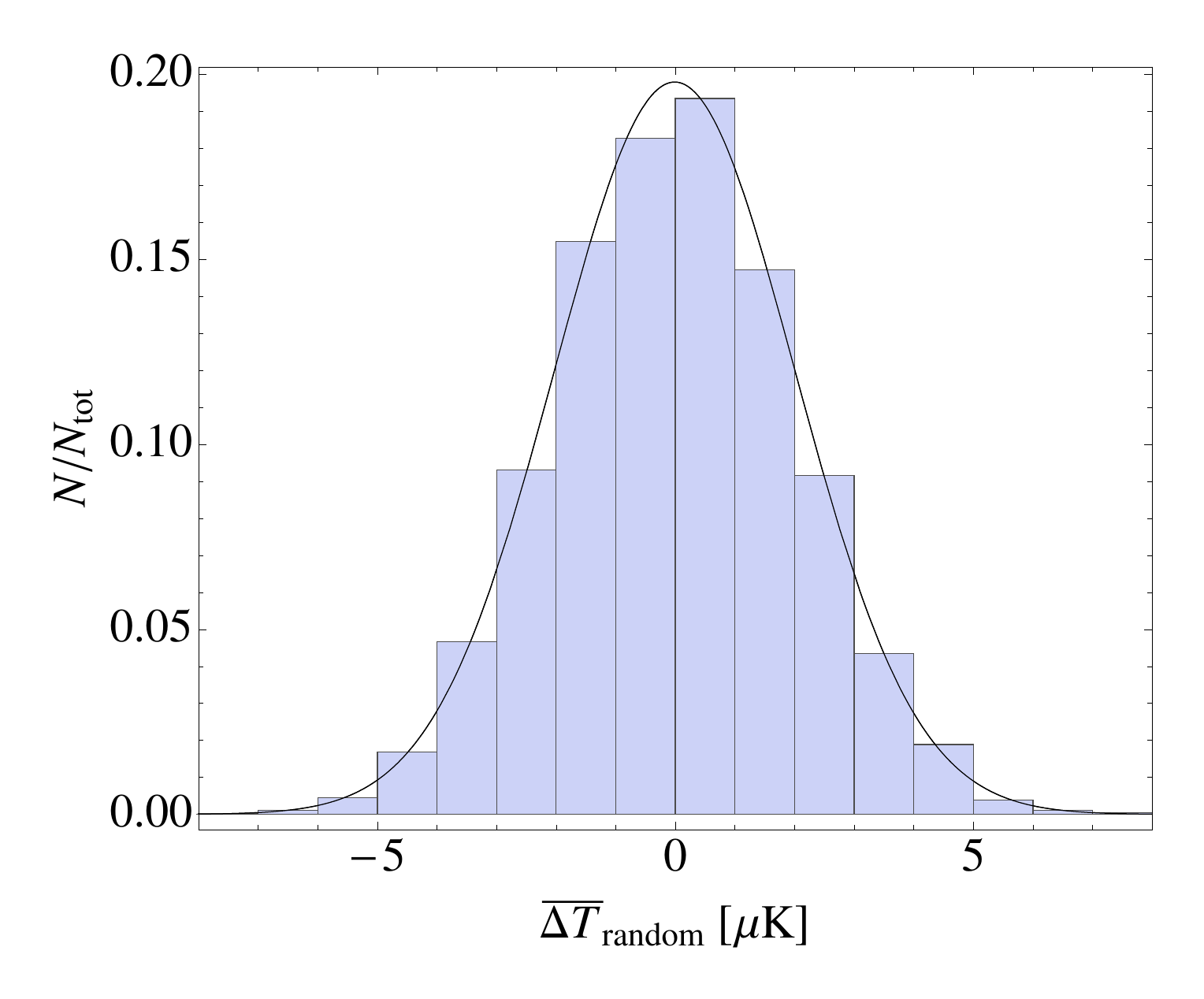}
\caption{\label{figure:histogram} 
An example histogram of the average $\overline{\Delta T}$ values obtained for $10000$ random selections of $50+50$ pixels on the filtered WMAP ILC map within a circular window of area $\sim8000$ square degrees. The filter radius is $\theta_F=4\dg$ and the average calculated as described in the text. The best-fit Gaussian distribution to the histogram is also plotted.}
\end{center}
\end{figure}

To account for the variation over the whole map, we repeat the measurement for $10$ different locations of the window and calculate the average $\overline{\Delta T}(N,\theta_F)$ and the variance $\sigma^2_{\Delta T}(N,\theta_F)$ over these locations for hot spots, cold spots and the total separately. The window locations are chosen to ensure coverage of the entire map. As a result there is some overlap between different windows and the $\Delta T$ values obtained from each are not completely independent. This overlap is increased if more windows are used, leading to an artificial underestimate of the variance. However, we checked that the mean $\overline{\Delta T}(N,\theta_F)$ remains stable. It should also be noted that each map is constructed from only one $1\;h^{-3}$Gpc$^3$ realisation of the density field, which needs to be used more than once to generate a map of the entire sky. Nevertheless, the mean $\overline{\Delta T}(N,\theta_F)$ obtained is an unbiased estimator of the true ensemble average value.

We emphasize again that $\overline{\Delta T}(N,\theta_F)$ obtained from the maps in this fashion is a measure of the maximum stacked  temperature signal that could be in principle be observed in a \lcdm cosmology by an ideal experiment in which the hot and cold spots were chosen based on an \emph{a posteriori} knowledge of the actual ISW temperature distribution. It should not be regarded as representative of the expected signal from any realistic observation. 

In order to assess the significance of the average maximum signal thus found, we need to estimate the noise introduced in the measurement by primary CMB fluctuations. To do this we use the WMAP 7-year Internal Linear Combination map\footnote{Available to download  from http://lambda.gsfc.nasa.gov/product/map/dr4/ilc\_map\_info.cfm.} and apply the compensated top-hat filter of radius $\theta_F$. We then choose $2N$ random pixels from within the same circular window as before, placed at a location chosen to maximize the overlap with the actual SDSS DR6 footprint, using the procedure described above to limit the amount of overlap. We add the temperatures of the first $N$ pixels, subtract the temperatures of the next $N$ pixels, and divide the result by $2N$ to obtain one random realization of $\Delta T$. We repeat this procedure $10,000$ times and fit a Gaussian distribution function to the resulting histogram of $\Delta T$. The normalized histogram and fitted distribution function for $N=50$ and $\theta_F=4\dg$ are shown in figure~\ref{figure:histogram}. The standard deviation of the fitted distribution provides the uncorrelated noise on the stacked ISW temperature measurement. The noise values we obtain agree well with those reported in the original measurement \cite{Granett:2008ju}.

\begin{figure}[t]
\begin{center}
\includegraphics[scale=0.5]{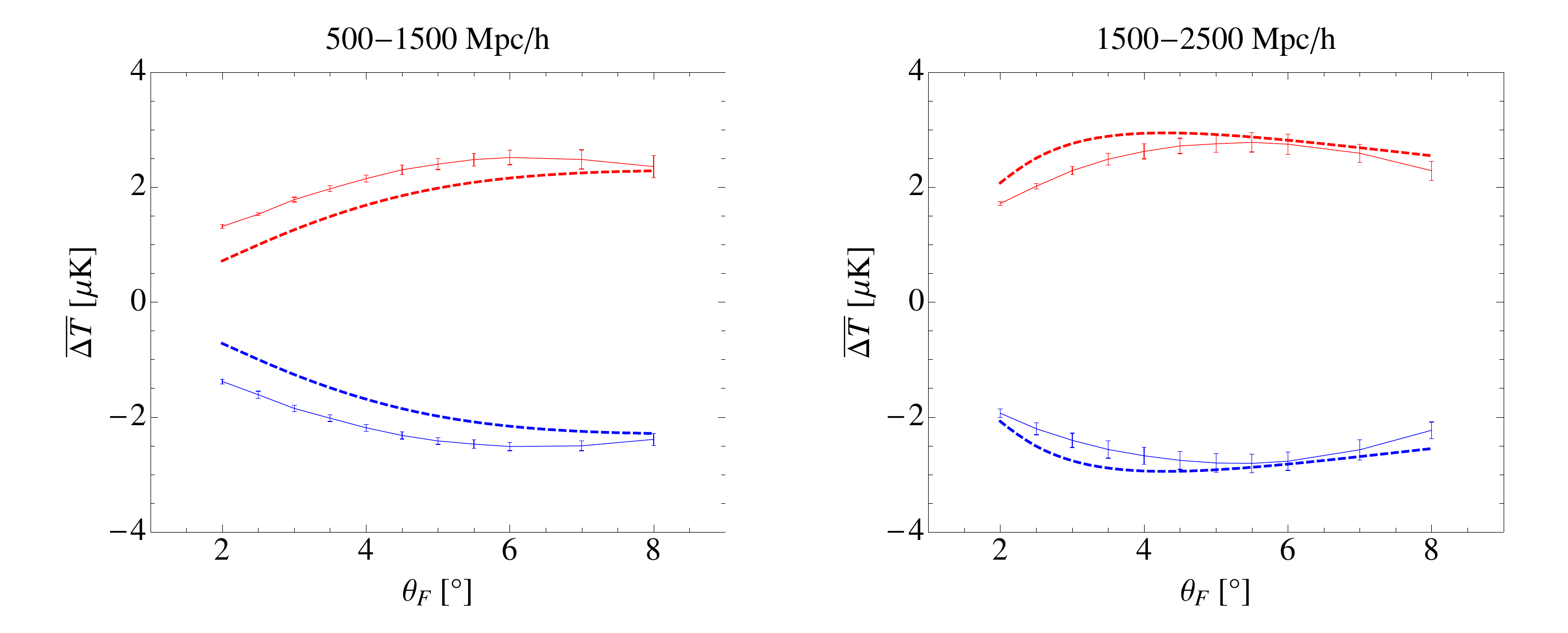}
\caption{\label{figure:TvsTheta} 
The solid curves show the variation of the maximum $\overline{\Delta T}_{\rm hot}$ (red) and $\overline{\Delta T}_{\rm cold}$ (blue) as a function of filter radius $\theta_F$ as determined from the simulated maps with $N=50$. 
The approximate error bars are obtained from the variation due to the choice of location of the window area on the maps. The dashed curves show the corresponding theoretical maximum values $\langle\Delta T_{\rm c}\rangle$ and $\langle\Delta T_{\rm v}\rangle$ obtained from the spherical model. The left panel is for Map A and the right panel for Map B.}
\end{center}
\end{figure}

\begin{figure}[t]
\begin{center}
\includegraphics[scale=0.55]{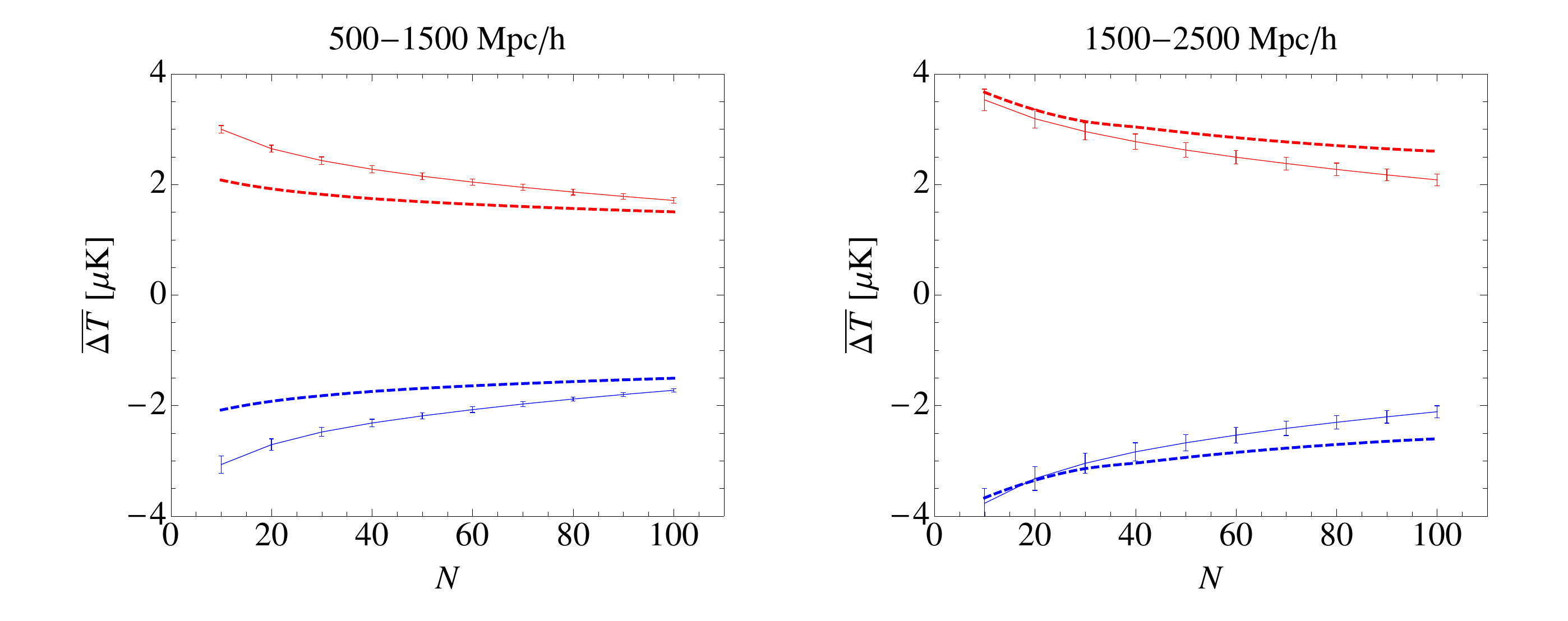}
\caption{\label{figure:TvsN} 
Same as figure \ref{figure:TvsTheta} except showing the variation of the maximum $\overline{\Delta T}$ as a function of $N$, for fixed filter radius $\theta_F=4^\circ$. }
\end{center}
\end{figure}

\section{Results and model comparison}
\label{section:results}

Figure~\ref{figure:TvsTheta} shows the variation of $\overline{\Delta T}(N,\theta_F)$ as a function of filter radius $\theta_F$ for $N=50$ obtained from the simulated maps for hot and cold spots separately. This represents the maximum possible ISW temperature signal in a $\Lambda$CDM universe. Also plotted is the theoretical prediction for the spherical model described in section~\ref{section:sphericalmodel}. The dependence of $\overline{\Delta T}(N,\theta_F)$ on $N$ for $\theta_F=4\dg$ is shown in figure~\ref{figure:TvsN}. 

Comparing the spherical model predictions and the upper limits $\overline{\Delta T}_{\rm hot/cold}$ obtained from Map A, which covers comoving distances in the range $500-1500\;h^{-1}$Mpc, we find that the spherical model somewhat underestimates the maximum possible signal that is in principle measurable by an ideal experiment. The values of $\overline{\Delta T}_{\rm hot}$ and $\overline{\Delta T}_{\rm cold}$ obtained from the maps are almost symmetrical, as would be expected for structures in the linear regime. This indicates that neither the non-linear evolution of density profiles nor the induced non-Gaussianity of the density distribution have any significant effect on the stacked ISW signal on the relevant scales.

Instead the reason for the difference in Map A is probably the assumption of sphericity. It is likely that the structures in the $N$-body simulation responsible for the very largest temperature shifts in this map are better described as elongated filamentary structures that happen to be aligned along the line of sight. By approximating them as spheres, the spherical model underestimates their ISW temperature shift. Since such long filaments that happen to be aligned along the line of sight are rare, their influence on the average $\overline{\Delta T}$ is relatively smaller when larger $N$ populations are considered, hence the amount by which the model prediction underestimates the actual maximal signal in Map A decreases at large $N$. The effect of neglecting correlations between extreme structures would also result in an artificial enhancement in the spherical model's prediction for large $N$.

It should be emphasized again that the spherical model prediction is for the expectation of the ISW temperature signal associated with peaks in the density fluctuation field, whereas the $\overline{\Delta T}_{\rm hot/cold}$ values are associated with peaks in the potential field and may not be directly observable. Therefore $\overline{\Delta T}_{\rm hot/cold}$ represent an upper bound on the maximum possible stacked ISW signal. Although for Map A the spherical model underestimates this maximum possible signal for the reasons stated above, the numerical difference is still small for all $N$ and $\theta_F$. In particular, at $N=50$ and $\theta_F=4\dg$ the difference is $\lesssim0.5\;\mu$K.

On the other hand for Map B, which covers comoving distances in the range $1500-2500\;h^{-1}$Mpc, the spherical model prediction for the maximum possible signal is slightly larger than the map values. At these distances, relatively narrow filamentary structures subtend a much smaller angle on the sky and so produce a smaller effective $\Delta T$ signal after application of the filter of width $\theta_F$. The structures that give the largest effective $\Delta T$ will be much wider, and therefore the spherical approximation is less important. Instead, effects such as the correlation between extreme structures mentioned above appear to be more relevant. As the spherical model \emph{overestimates} the maximum possible signal at these high redshifts it can be used to derive a conservative upper bound on the observable signal. 

\begin{figure}[t]
\begin{center}
\includegraphics[scale=0.58]{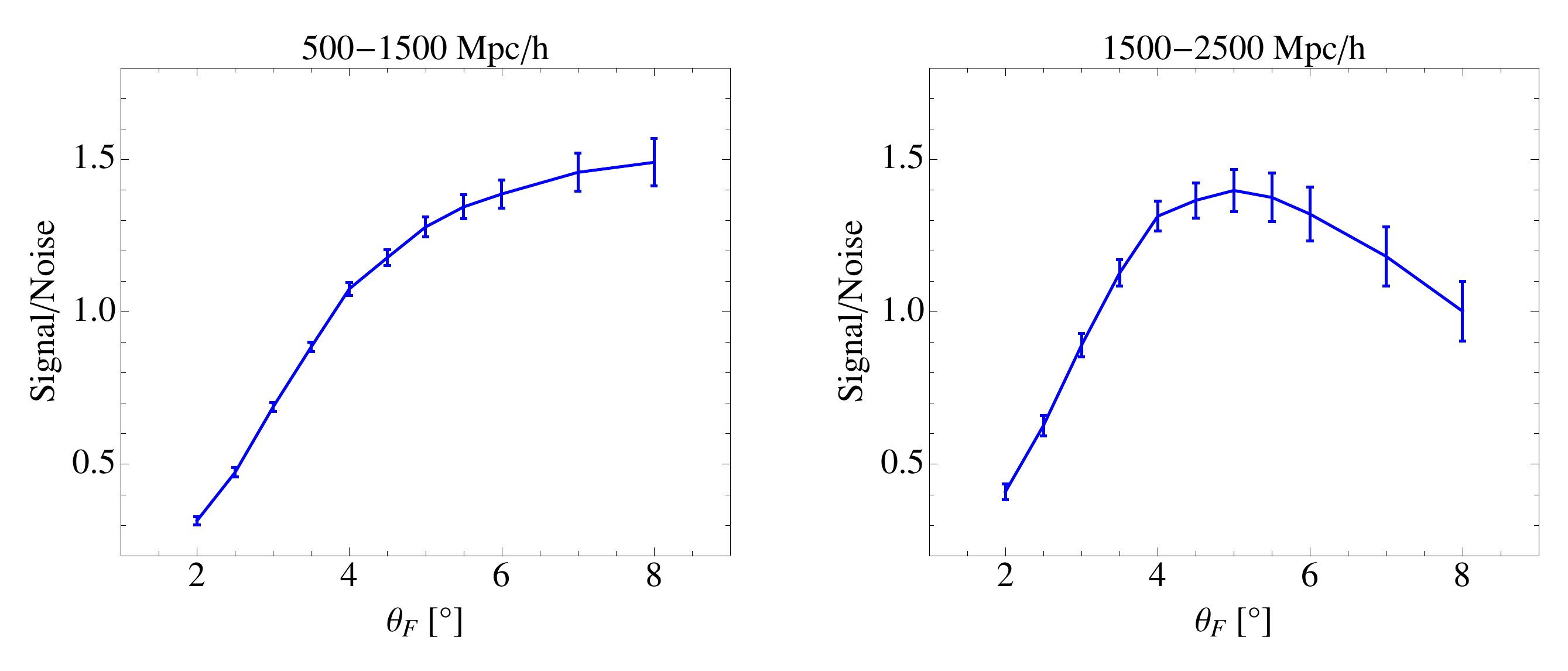}
\caption{\label{figure:SvN1} 
The variation of the maximum achievable signal-to-noise ratio as a function of filter radius $\theta_F$, for fixed $N=50$. The maximum signal is calculated by subtracting the average of the cold spots from the average of hot spots and dividing by $2N$, and the errors are added in quadrature. The noise is calculated from the width of the Gaussian fit to the distribution of $\overline{\Delta T}$ values found for a random location of patches on the WMAP ILC map, as described in detail in the text.}
\end{center}
\end{figure}

\begin{figure}[t]
\begin{center}
\includegraphics[scale=0.58]{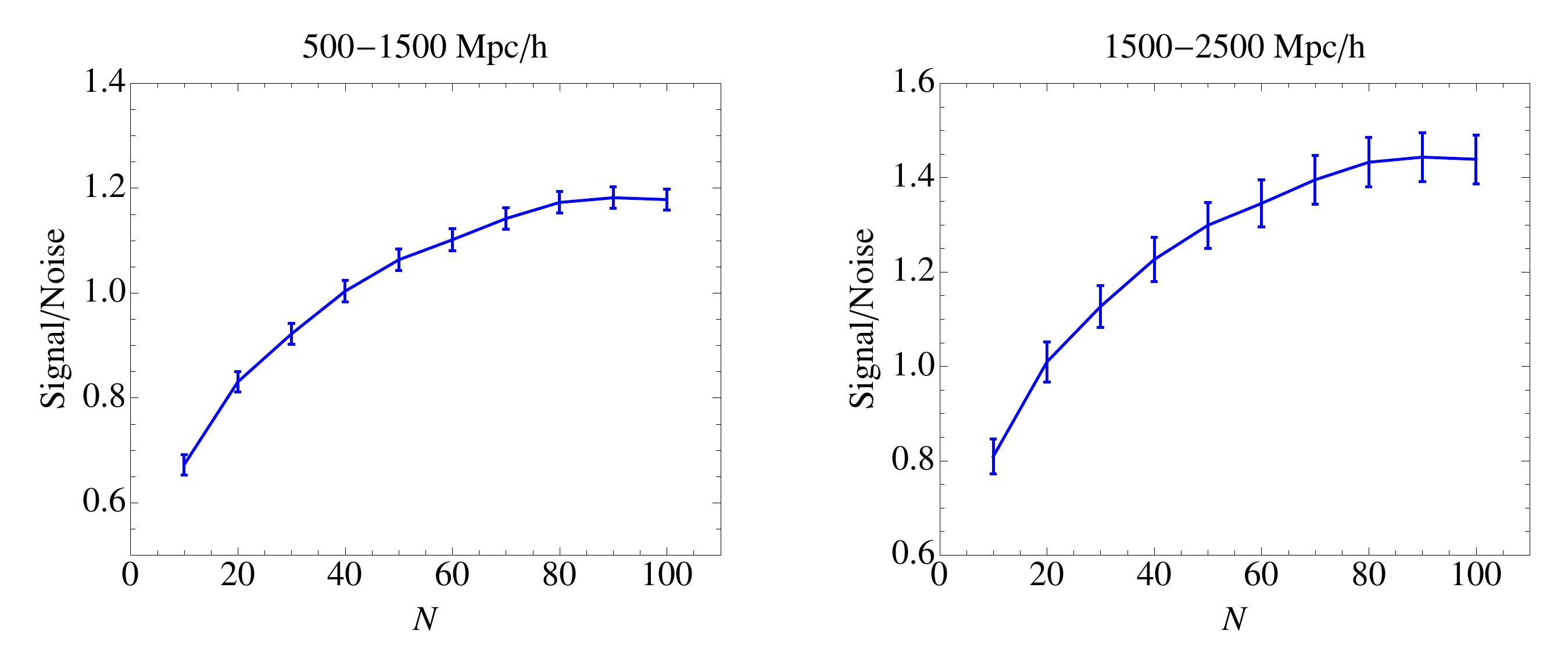}
\caption{\label{figure:SvN2} 
Same as figure \ref{figure:SvN1}, except showing the maximum achievable signal-to-noise ratio as a function of $N$, for fixed $\theta_F=4^\circ$.}
\end{center}
\end{figure}

Another quantity of interest is the ratio of the maximum possible stacked signal obtained from the map analysis to the intrinsic observational noise in the measurement due to primary CMB anisotropies. In figures~\ref{figure:SvN1} and \ref{figure:SvN2} we plot the behaviour of the signal-to-noise ratio as a function of $N$ and $\theta_F$. In this case while varying $\theta_F$ the pixel selection is always based on the map filtered with the $4\dg$ filter in order to keep the same sample of structures. The signal-to-noise ratio does not show a peak at $N=50$ or at $\theta_F=4\dg$ for either of the maps analyzed, in contrast to the observation in ref.~\cite{Granett:2008ju}. It is also clear that even though by construction the map analysis selects precisely the regions contributing the largest filtered signal, the maximum signal-to-noise ratio in a $\Lambda$CDM cosmology with Gaussian perturbations is always less than $1.5$. A realistic observation would not even be able to achieve this value. We conclude that the observed $4.4\sigma$ significance of the signal is totally at odds with the concordance cosmology.

\section{Conclusions}
\label{section:conclusions}

The results of the previous section show that at high redshifts, the spherical model calculation for the maximum possible stacked ISW signal that could be observed is an overestimate of the actual value obtained from the simulated map as described. At lower redshifts the \emph{a posteriori} map analysis gives a larger maximum signal than the spherical model calculation, though the quantitative differences are small.

The symmetry between hot and cold spots in the simulated maps shows that the linear treatment in the spherical model is valid and that the non-Gaussianity induced by gravitational evolution does not produce a significant effect.

Instead at low redshifts the spherical model underestimates the $\overline{\Delta T}$ from the simulated map primarily due to the assumption of sphericity. Whether the actual maximum possible stacked ISW signal in a realistic observation is closer to the spherical model prediction or the value from the map analysis depends on the properties of the galaxy catalogue and the strategy that is used to identify the superstructures. For a catalogue with spectroscopically determined redshifts, the redshift smearing problem described in section~\ref{section:limitations} is less severe. Therefore a structure-finding algorithm that does not have a directional preference will select a population of structures that, although not spherical, will have uniformly distributed orientations and whose mean effect can therefore be well described by the spherical model. On the other hand when only photometric redshifts are available, the redshift smearing induced will naturally bias the population of structures selected towards including more long structures aligned along the observer's line of sight. In such a case despite its conservative assumptions, the spherical model underestimates the maximum stacked signal that can in principle be observed.\footnote{We note in passing that in the presence of such redshift errors, searching for structures in the reconstructed 3D density field is likely not the optimal strategy in any case.} 

Of course the closeness of the two predictions at low redshifts mean that this is a moot point. The values of $\langle\Delta T\rangle$ and $\overline{\Delta T}$ obtained from Map A for the redshift range $0.17<z<0.57$ differ by less than $0.5\;\mu$K at $N=50$ and $\theta_F=4\dg$. For Map B, over the redshift range $0.57<z<1.08$, the difference is even smaller and the spherical model \emph{overestimates} the obtainable signal.

We therefore conclude that the spherical model prediction $\langle\Delta T\rangle=2.27\pm0.14\;\mu$K using the actual SDSS DR6 survey volume and redshift distribution can meaningfully be compared with the observed value of $\Delta T=9.6\pm2.2\;\mu$K for the combined signal of voids and clusters~\cite{Granett:2008ju}. The observed signal is easily $>3\sigma$ larger than even the most optimistic expectation in the standard $\Lambda$CDM cosmological model with Gaussian primordial perturbations. This confirms the conclusion reached in ref.~\cite{Nadathur:2011iu} that the observed correlation cannot be due to the linear ISW effect in such a cosmology.

The question that then arises is, what does account for this large correlation between structures in the LRG sample and CMB temperature fluctuations? The immediate assumption is that it is a spurious signal caused by some as yet undetected systematic or foreground contamination in the observation. It is however difficult to explain how this might be the case. Contamination in the CMB maps from extragalactic radio sources or due to Sunyaev-Zel'dovich (SZ) effect is potentially possible. However, the original observation method used CMB maps with the conservative KQ75 mask applied and foreground templates subtracted to avoid contamination from the Galaxy and point sources. The authors also checked for, but did not find, any frequency dependence of the signal over the Q, V and W band CMB maps in the absence of foreground subtraction \cite{Granett:2008ju}. This suggests that foreground contamination is not an issue. 

The SZ effect would be expected to reduce the signal strength for the hot spots associated with overdensities, though the few clusters present in voids might enhance the signal seen for cold spots. The characteristic scale of this effect is however small compared to the filter radius used here, and so the net effect would be small. Nevertheless, it would be a worthwhile exercise to repeat the same observational method using a mask from the improved SZ point source catalogue available from Planck.

If foreground contamination is not important and no other systematic effect can be found, it is still possible that the observation is simply due to an unfortunate correlation between primary CMB fluctuations and the directions along which superstructures were identified in the LRG catalogue. Such a correlation is unlikely at a $>3\sigma$ level given the size of the ISW signal expected in $\Lambda$CDM (if selecting CMB patches at random without any reference to large-scale structure it is $>4.4\sigma$ unlikely), but even such rare fluctuations do of course sometimes occur. However, if the signal is caused simply by such an unlucky fluctuation, it should disappear if the same observation is repeated with another catalogue of galaxies or other tracers of the matter density field that covers a different region of the sky. Efforts to test for such a correlation are somewhat complicated by the fact that in the absence of accurate redshift information for the given catalogue, the method used here to locate structures in the 3D distribution is not optimal and other techniques must be devised. This is work in progress.

If such further analyses should confirm the existence of the stacked ISW signal then some cosmological explanation is required. This would be a strong indication of the need for some new physics beyond the standard $\Lambda$CDM model. We have previously speculated as to what such new physics might be \cite{Nadathur:2011iu}. Taken at face value, the observation indicates that the rarest and most extreme fluctuations in the matter density contrast field $\delta$ are in fact more numerous than expected. This means that the extreme tails of the probability distribution function for $\delta$ are modified from the Gaussian expectation on linear scales. Such a modification could arise from a primordial non-Gaussian kurtosis in the matter density field, though the magnitude of the kurtosis required to explain the results is still to be calculated. Alternatively, the explanation could lie in the late-time growth of structures in an alternative to $\Lambda$CDM, such as modified gravity or $f(R)$ models.

Such a modification of the Gaussian distribution function of $\delta$ would also affect the full cross-correlation amplitude for the ISW effect measured in other studies, although such observations are not specifically designed to test the tail of the distribution so might be less sensitive to deviations. We note again that mild excesses of the full cross-correlation amplitude have sometimes been found when using the same LRG catalogues as tracers \cite{Giannantonio:2008zi,Ho:2008bz,Giannantonio:2012aa,Granett:2008dz}.

We conclude by noting that the stacked ISW signal of rare superstructures is one of the most statistically significant anomalies that has yet been found in the standard cosmological model.  The ISW effect is a linear order effect and in this work we have demonstrated that the theoretical calculation of the expectation value on relevant scales is not affected by complexities of non-linear dynamics. It is also independent of any assumptions about the precise nature of the bias relationship between the galaxy distribution and the matter density field. This may be taken as further reason to believe in the robustness of this anomaly.
 
\acknowledgments
We are grateful to Yan-Chuan Cai for providing us with the maps used in this analysis, and we thank Steve Maddox for helpful discussions. SH acknowledges support from the Academy of Finland grant 131454. SN acknowledges support from the Deutsche Forschungsgemeinschaft (DFG) via the International Research Training Group 881 `Quantum Fields and Strongly Interacting Matter', and from the Sofja Kovalevskaja program of the Alexander von Humboldt Foundation. 

\paragraph{Note added:} Shortly after submission of this paper, an independent study of the same issue was uploaded to the arXiv~\cite{HernandezMonteagudo:2012ms}. This study uses direct $N$-body simulations of the relevant redshift range and finds the same signal size for $\Lambda$CDM as presented here. The authors also re-perform the stacking analysis and confirm that the signal seen in \cite{Granett:2008ju} is robust to rotation tests of the CMB maps, while finding no evidence for foreground contamination. 

\appendix
\section{Statistics of Gaussian random fields}
\label{appendix}
In this appendix, we provide definitions for some of the quantities and functions used in section \ref{section:sphericalmodel}. A full derivation of these expressions from a treatment of the statistics of Gaussian random fields is provided in ref.~\cite{BBKS}; here we limit ourselves to the details required for completeness of this paper.

Consider a homogeneous and isotropic Gaussian random field $\delta(\mathbf{r})$, with two-point correlation function $\xi(|\mathbf{r}_1-\mathbf{r}_2|)=\langle\delta(\mathbf{r}_1)\delta(\mathbf{r}_2)\rangle$ and power spectrum $P(k)$ defined as the Fourier transform of $\xi(r)$. The field is smoothed on length scale $R_{\rm f}$ using a Gaussian filter, such that in Fourier space the smoothed field and its power spectrum are
\beq
\delta_{\rm f}(k) = \exp\left(-R_{\rm f}^2k^2/2\right)\delta(k)\,,~~~P_{\rm f}(k)=\exp(-R_{\rm f}^2k^2)P(k)\;.
\eeq
The spectral moments of this field, $\sigma_j$, are defined as
\beq
\label{eq:sigma}
\sigma_j^2\equiv\int_0^\infty\frac{k^2{\rm d}k}{2\pi^2}P_{\rm f}(k)k^{2j}\;,
\eeq
and the corresponding spectral parameters are
\beq
\label{eq:specparams}
\gamma\equiv\frac{\sigma_1^2}{\sigma_2\sigma_0}\,,~~~~R_\ast\equiv\sqrt{3}\frac{\sigma_1}{\sigma_2}\;.
\eeq
Throughout the paper for the purposes of numerical evaluation we take a smoothing scale of $R_{\rm f}=20\;h^{-1}$Mpc.

The expression for the differential comoving number density of peaks of the smoothed field as a function of $\nu$ and $x$ can then be expressed in terms of these spectral parameters as in eq.~\eqref{eq:numberdensity}. The function $f(x)$ appearing in this expression can be written in closed form as 
\begin{eqnarray}
\label{eq:fofx}
f(x)=\frac{(x^3-3x)}{2}&&\left\{{\rm erf}\left[\left(\frac{5}{2}\right)^{1/2}x\right]+{\rm erf}\left[\left(\frac{5}{2}\right)^{1/2}\frac{x}{2}\right]\right\} \nonumber \\ 
&&+\left(\frac{2}{5\pi}\right)^{1/2}\left[\left(\frac{31x^2}{4}+\frac{8}{5}\right)e^{-5x^2/8}+\left(\frac{x^2}{2}-\frac{8}{5}\right)e^{-5x^2/2}\right]\,.
\end{eqnarray}
The mean spherical profile about a peak with given $\nu$ and $x$ can be written as
\beq
\label{eq:BBKSdensityprofile}
\bar\delta_{\rm pk}(r;\nu,x) = \frac{\nu\sigma_0}{(1-\gamma^2)}\left(\psi+\frac{R_\ast^2}{3}\nabla^2\psi\right)-\frac{x\sigma_0}{(1-\gamma^2)}\left(\gamma^2\psi+\frac{R_\ast^2}{3}\nabla^2\psi\right)\,
\eeq
where $\psi(r)\equiv\xi(r)/\xi(0)$.~Writing $\xi(r)$ in terms of its Fourier transform, using $\xi(0)=\sigma_0^2$ and rearranging terms leads to eq.~\eqref{eq:profile}.

\bibliography{isw.bib}

\providecommand{\href}[2]{#2}\begingroup\raggedright\begin{thebibliography}{10}

\bibitem{Crittenden:1995ak}
R.~G. Crittenden and N.~Turok, {\it {Looking for Lambda with the Rees-Sciama
  effect}},  {\em Phys.Rev.Lett.} {\bf 76} (1996) 575,
  [\href{http://xxx.lanl.gov/abs/astro-ph/9510072}{{\tt astro-ph/9510072}}].

\bibitem{Sachs:1967er}
R.~Sachs and A.~Wolfe, {\it {Perturbations of a cosmological model and angular
  variations of the microwave background}},  {\em Astrophys.J.} {\bf 147}
  (1967) 73--90.

\bibitem{Afshordi:2004kz}
N.~Afshordi, {\it {Integrated Sachs-Wolfe effect in cross - correlation: The
  Observer's manual}},  {\em Phys.Rev.} {\bf D70} (2004) 083536,
  [\href{http://xxx.lanl.gov/abs/astro-ph/0401166}{{\tt astro-ph/0401166}}].

\bibitem{Douspis:2008xv}
M.~Douspis, P.~G. Castro, C.~Caprini, and N.~Aghanim, {\it {Optimising large
  galaxy surveys for ISW detection}},  {\em Astron.Astrophys.} {\bf 485} (2008)
  395, [\href{http://xxx.lanl.gov/abs/0802.0983}{{\tt arXiv:0802.0983}}].

\bibitem{Boughn:2003yz}
S.~Boughn and R.~Crittenden, {\it {A correlation of the cosmic microwave sky
  with large scale structure}},  {\em Nature} {\bf 427} (2004) 45--47,
  [\href{http://xxx.lanl.gov/abs/astro-ph/0305001}{{\tt astro-ph/0305001}}].

\bibitem{Afshordi:2003xu}
N.~Afshordi, Y.-S. Loh, and M.~A. Strauss, {\it {Cross-correlation of the
  Cosmic Microwave Background with the 2MASS galaxy survey: signatures of dark
  energy, hot gas, and point sources}},  {\em Phys.Rev.} {\bf D69} (2004)
  083524, [\href{http://xxx.lanl.gov/abs/astro-ph/0308260}{{\tt
  astro-ph/0308260}}].

\bibitem{Fosalba:2003ge}
P.~Fosalba, E.~Gaztanaga, and F.~Castander, {\it {Detection of the ISW and SZ
  effects from the CMB-galaxy correlation}},  {\em Astrophys.J.} {\bf 597}
  (2003) L89--92, [\href{http://xxx.lanl.gov/abs/astro-ph/0307249}{{\tt
  astro-ph/0307249}}].

\bibitem{Nolta:2003uy}
WMAP Collaboration, M.~R. Nolta {\em et.~al.}, {\it {First year Wilkinson
  Microwave Anisotropy Probe (WMAP) observations: Dark energy induced
  correlation with radio sources}},  {\em Astrophys.J.} {\bf 608} (2004)
  10--15, [\href{http://xxx.lanl.gov/abs/astro-ph/0305097}{{\tt
  astro-ph/0305097}}].

\bibitem{Padmanabhan:2004fy}
N.~Padmanabhan, C.~M. Hirata, U.~Seljak, D.~Schlegel, J.~Brinkmann, {\em
  et.~al.}, {\it {Correlating the CMB with luminous red galaxies: The
  Integrated Sachs-Wolfe effect}},  {\em Phys.Rev.} {\bf D72} (2005) 043525,
  [\href{http://xxx.lanl.gov/abs/astro-ph/0410360}{{\tt astro-ph/0410360}}].

\bibitem{Cabre:2006qm}
A.~Cabre, E.~Gaztanaga, M.~Manera, P.~Fosalba, and F.~Castander, {\it
  {Cross-correlation of WMAP 3rd year and the SDSS DR4 galaxy survey: new
  evidence for dark energy}},  {\em Mon.Not.Roy.Astron.Soc.} {\bf 372} (2006)
  L23--L27, [\href{http://xxx.lanl.gov/abs/astro-ph/0603690}{{\tt
  astro-ph/0603690}}].

\bibitem{Rassat:2006kq}
A.~Rassat, K.~Land, O.~Lahav, and F.~B. Abdalla, {\it {Cross-correlation of
  2MASS and WMAP3: Implications for the Integrated Sachs-Wolfe effect}},  {\em
  Mon.Not.Roy.Astron.Soc.} {\bf 377} (2007) 1085--1094,
  [\href{http://xxx.lanl.gov/abs/astro-ph/0610911}{{\tt astro-ph/0610911}}].

\bibitem{Giannantonio:2006du}
T.~Giannantonio, R.~G. Crittenden, R.~C. Nichol, R.~Scranton, G.~T. Richards,
  {\em et.~al.}, {\it {A high redshift detection of the integrated Sachs-Wolfe
  effect}},  {\em Phys.Rev.} {\bf D74} (2006) 063520,
  [\href{http://xxx.lanl.gov/abs/astro-ph/0607572}{{\tt astro-ph/0607572}}].

\bibitem{Raccanelli:2008bc}
A.~Raccanelli, A.~Bonaldi, M.~Negrello, S.~Matarrese, G.~Tormen, {\em et.~al.},
  {\it {A reassessment of the evidence of the integrated Sachs-Wolfe effect
  through the WMAP-NVSS correlation}},  {\em Mon.Not.Roy.Astron.Soc.} {\bf 386}
  (2008) 2161, [\href{http://xxx.lanl.gov/abs/0802.0084}{{\tt
  arXiv:0802.0084}}].

\bibitem{HernandezMonteagudo:2009fb}
C.~Hernandez-Monteagudo, {\it {Revisiting the WMAP - NVSS angular cross
  correlation. A skeptic view}},  \href{http://xxx.lanl.gov/abs/0909.4294}{{\tt
  arXiv:0909.4294}}.

\bibitem{Francis:2009ps}
C.~Francis and J.~Peacock, {\it {ISW measurements with photometric redshift
  surveys: 2MASS results and future prospects}},  {\em Mon.Not.Roy.Astron.Soc.}
  {\bf 406} (2010) 2, [\href{http://xxx.lanl.gov/abs/0909.2494}{{\tt
  arXiv:0909.2494}}].

\bibitem{Sawangwit:2009gd}
U.~Sawangwit, T.~Shanks, R.~Cannon, S.~Croom, N.~P. Ross, {\em et.~al.}, {\it
  {Cross-correlating WMAP5 with 1.5 million LRGs: a new test for the ISW
  effect}},  {\em Mon.Not.Roy.Astron.Soc.} {\bf 402} (2010) 2228,
  [\href{http://xxx.lanl.gov/abs/0911.1352}{{\tt arXiv:0911.1352}}].

\bibitem{LopezCorredoira:2010rr}
M.~Lopez-Corredoira, F.~S. Labini, and J.~Betancort-Rijo, {\it {Absence of
  significant cross-correlation between WMAP and SDSS}},  {\em
  Astron.Astrophys.} {\bf 513} (2010) A3,
  [\href{http://xxx.lanl.gov/abs/1001.4000}{{\tt arXiv:1001.4000}}].

\bibitem{Dupe:2010zs}
F.-X. Dupe, A.~Rassat, J.-L. Starck, and M.~Fadili, {\it {Measuring the
  integrated Sachs-Wolfe Effect}},  {\em Astron.Astrophys.} {\bf 534} (2011)
  A51, [\href{http://xxx.lanl.gov/abs/1010.2192}{{\tt arXiv:1010.2192}}].

\bibitem{Giannantonio:2008zi}
T.~Giannantonio, R.~Scranton, R.~G. Crittenden, R.~C. Nichol, S.~P. Boughn,
  {\em et.~al.}, {\it {Combined analysis of the integrated Sachs-Wolfe effect
  and cosmological implications}},  {\em Phys.Rev.} {\bf D77} (2008) 123520,
  [\href{http://xxx.lanl.gov/abs/0801.4380}{{\tt arXiv:0801.4380}}].

\bibitem{Ho:2008bz}
S.~Ho, C.~Hirata, N.~Padmanabhan, U.~Seljak, and N.~Bahcall, {\it {Correlation
  of CMB with large-scale structure: I. ISW Tomography and Cosmological
  Implications}},  {\em Phys.Rev.} {\bf D78} (2008) 043519,
  [\href{http://xxx.lanl.gov/abs/0801.0642}{{\tt arXiv:0801.0642}}].

\bibitem{Giannantonio:2012aa}
T.~Giannantonio, R.~Crittenden, R.~Nichol, and A.~J. Ross, {\it {The
  significance of the integrated Sachs-Wolfe effect revisited}},  {\em
  Mon.Not.Roy.Astron.Soc.} {\bf 426} (2012) 2581,
  [\href{http://xxx.lanl.gov/abs/1209.2125}{{\tt arXiv:1209.2125}}].

\bibitem{Granett:2008dz}
B.~R. Granett, M.~C. Neyrinck, and I.~Szapudi, {\it {A Map of the Integrated
  Sachs-Wolfe Signal from Luminous Red Galaxies}},  {\em Astrophys.J.} {\bf
  701} (2009) 414--422, [\href{http://xxx.lanl.gov/abs/0812.1025}{{\tt
  arXiv:0812.1025}}].

\bibitem{Goto:2012}
T.~Goto, I.~Szapudi, and B.~R. Granett, {\it {Cross-correlation of WISE
  galaxies with the cosmic microwave background}},  {\em
  Mon.Not.Roy.Astron.Soc.} {\bf 422} (2012), no.~1 L77--L81,
  [\href{http://xxx.lanl.gov/abs/1202.5306}{{\tt arXiv:1202.5306}}].

\bibitem{Granett:2008ju}
B.~R. Granett, M.~C. Neyrinck, and I.~Szapudi, {\it {An Imprint of
  Super-Structures on the Microwave Background due to the Integrated
  Sachs-Wolfe Effect}},  {\em Astrophys.J.} {\bf 683} (2008) L99,
  [\href{http://xxx.lanl.gov/abs/0805.3695}{{\tt arXiv:0805.3695}}].

\bibitem{Granett:2008xb}
B.~R. Granett, M.~C. Neyrinck, and I.~Szapudi, {\it {Dark Energy Detected with
  Supervoids and Superclusters}},
  \href{http://xxx.lanl.gov/abs/0805.2974}{{\tt arXiv:0805.2974}}.

\bibitem{Hunt:2008wp}
P.~Hunt and S.~Sarkar, {\it {Constraints on large scale inhomogeneities from
  WMAP-5 and SDSS: confrontation with recent observations}},  {\em
  Mon.Not.Roy.Astron.Soc.} {\bf 401} (2010) 547,
  [\href{http://xxx.lanl.gov/abs/0807.4508}{{\tt arXiv:0807.4508}}].

\bibitem{Inoue:2010rp}
K.~T. Inoue, N.~Sakai, and K.~Tomita, {\it {Evidence of Quasi-linear
  Super-Structures in the Cosmic Microwave Background and Galaxy
  Distribution}},  {\em Astrophys.J.} {\bf 724} (2010) 12--25,
  [\href{http://xxx.lanl.gov/abs/1005.4250}{{\tt arXiv:1005.4250}}].

\bibitem{Nadathur:2011iu}
S.~Nadathur, S.~Hotchkiss, and S.~Sarkar, {\it {The integrated Sachs-Wolfe
  imprints of cosmic superstructures: a problem for $\Lambda$CDM}},  {\em JCAP}
  {\bf 1206} (2012) 042, [\href{http://xxx.lanl.gov/abs/1109.4126}{{\tt
  arXiv:1109.4126}}].

\bibitem{Cai:2010hx}
Y.-C. Cai, S.~Cole, A.~Jenkins, and C.~S. Frenk, {\it {Full-sky map of the ISW
  and Rees-Sciama effect from Gpc simulations}},
  \href{http://xxx.lanl.gov/abs/1003.0974}{{\tt arXiv:1003.0974}}.

\bibitem{AdelmanMcCarthy:2007aa}
SDSS Collaboration, J.~K. Adelman-McCarthy {\em et.~al.}, {\it {The Sixth Data
  Release of the Sloan Digital Sky Survey}},  {\em Astrophys.J.Suppl.} {\bf
  175} (2008) 297--313, [\href{http://xxx.lanl.gov/abs/0707.3413}{{\tt
  arXiv:0707.3413}}].

\bibitem{Neyrinck:2004gj}
M.~C. Neyrinck, N.~Y. Gnedin, and A.~J. Hamilton, {\it {VOBOZ: an
  almost-parameter-free halo-finding algorithm}},  {\em
  Mon.Not.Roy.Astron.Soc.} {\bf 356} (2005) 1222,
  [\href{http://xxx.lanl.gov/abs/astro-ph/0402346}{{\tt astro-ph/0402346}}].

\bibitem{Neyrinck:2007gy}
M.~C. Neyrinck, {\it {ZOBOV: a parameter-free void-finding algorithm}},
  \href{http://xxx.lanl.gov/abs/0712.3049}{{\tt arXiv:0712.3049}}.

\bibitem{Hinshaw:2008kr}
WMAP Collaboration, G.~Hinshaw {\em et.~al.}, {\it {Five-Year Wilkinson
  Microwave Anisotropy Probe (WMAP) Observations: Data Processing, Sky Maps,
  and Basic Results}},  {\em Astrophys.J.Suppl.} {\bf 180} (2009) 225--245,
  [\href{http://xxx.lanl.gov/abs/0803.0732}{{\tt arXiv:0803.0732}}].

\bibitem{BBKS}
J.~M. Bardeen, J.~Bond, N.~Kaiser, and A.~Szalay, {\it {The Statistics of Peaks
  of Gaussian Random Fields}},  {\em Astrophys.J.} {\bf 304} (1986) 15--61.

\bibitem{Lahav:1991}
O.~Lahav and P.~B. Lilje, {\it {Evolution of velocity and density fields around
  clusters of galaxies}},  {\em Astrophys.J.} {\bf 374} (1991) 29--43.

\bibitem{Komatsu:2010fb}
WMAP Collaboration, E.~Komatsu {\em et.~al.}, {\it {Seven-Year Wilkinson
  Microwave Anisotropy Probe (WMAP) Observations: Cosmological
  Interpretation}},  {\em Astrophys.J.Suppl.} {\bf 192} (2011) 18,
  [\href{http://xxx.lanl.gov/abs/1001.4538}{{\tt arXiv:1001.4538}}].

\bibitem{Abazajian:2008wr}
SDSS Collaboration, K.~N. Abazajian {\em et.~al.}, {\it {The Seventh Data
  Release of the Sloan Digital Sky Survey}},  {\em Astrophys.J.Suppl.} {\bf
  182} (2009) 543--558, [\href{http://xxx.lanl.gov/abs/0812.0649}{{\tt
  arXiv:0812.0649}}].

\bibitem{Lewis:1999bs}
A.~Lewis, A.~Challinor, and A.~Lasenby, {\it {Efficient computation of CMB
  anisotropies in closed FRW models}},  {\em Astrophys.J.} {\bf 538} (2000)
  473--476, [\href{http://xxx.lanl.gov/abs/astro-ph/9911177}{{\tt
  astro-ph/9911177}}].

\bibitem{Rees:1968zza}
M.~Rees and D.~Sciama, {\it {Large scale Density Inhomogeneities in the
  Universe}},  {\em Nature} {\bf 217} (1968) 511--516.

\bibitem{Gorski:2004by}
K.~Gorski, E.~Hivon, A.~Banday, B.~Wandelt, F.~Hansen, {\em et.~al.}, {\it
  {HEALPix - A Framework for high resolution discretization, and fast analysis
  of data distributed on the sphere}},  {\em Astrophys.J.} {\bf 622} (2005)
  759--771, [\href{http://xxx.lanl.gov/abs/astro-ph/0409513}{{\tt
  astro-ph/0409513}}].

\bibitem{HernandezMonteagudo:2012ms}
C.~Hernandez-Monteagudo and R.~E. Smith, {\it {On the signature of nearby
  superclusters and voids in the Integrated Sachs-Wolfe effect}},
  \href{http://xxx.lanl.gov/abs/1212.1174}{{\tt arXiv:1212.1174}}.

\end{thebibliography}\endgroup
\bibliographystyle{JHEP}

\end{document}